# NEAR EARTH OBJECT 2012XJ112 AS A SOURCE OF BRIGHT BOLIDES OF ACHONDRITIC NATURE

**José M. Madiedo[1,2], Josep M. Trigo-Rodríguez[3], Iwan P. Williams[4], Natalia Konovalova[5], José L. Ortiz[6], Alberto J. Castro-Tirado[6], Sensi Pastor[7], José A. de los Reyes[7] and Jesús Cabrera-Caño[2]**

[1] Facultad de Ciencias Experimentales, Universidad de Huelva. 21071 Huelva (Spain).

[2] Departamento de Física Atómica, Molecular y Nuclear. Facultad de Física. Universidad de Sevilla. 41012 Sevilla, Spain.

[3] Institute of Space Sciences (CSIC-IEEC), Campus UAB, Facultat de Ciències, Torre C5-parell-2ª, 08193 Bellaterra, Barcelona, Spain.

[4] School of Physics and Astronomy, Queen Mary, University of London, London, UK.

[5] Institute of Astrophysics of the Academy of Sciences of the Republic of Tajikistan, Bukhoro, str. 22, Dushanbe 734042, Tajikistan.

[6] Instituto de Astrofísica de Andalucía, CSIC, Apt. 3004, Camino Bajo de Huetor 50, 18080 Granada, Spain.

[7] Observatorio Astronómico de La Murta. Molina de Segura, 30500 Murcia, Spain.

**SHORT TITLE:** NEO 2012XJ112 AS A SOURCE OF BRIGHT BOLIDES

**ABSTRACT**

We analyze the likely link between the recently discovered Near Earth Object 2012XJ112 and a bright fireball observed over the south of Spain on December 27, 2012. The bolide, with an absolute magnitude of -9±1, was simultaneously imaged during the morning twilight from two meteor stations operated by the SPanish Meteor Network (SPMN). It was also observed by several casual witnesses. The emission spectrum produced during the ablation of the meteoroid in the atmosphere was also recorded. From its analysis the chemical nature of this particle was inferred. Although our orbital association software identified several potential parent bodies for this meteoroid, the analysis of the evolution of the orbital elements performed with the Mercury 6 symplectic integrator supports the idea that NEO 2012XJ112 is the source of this meteoroid. The implications of this potential association are discussed here. In particular, the meteoroid bulk chemistry is consistent with a basaltic achondrite, and this emphasizes the importance to deduce from future Earth approaches the reflectance spectrum and taxonomic nature of 2012XJ112.

# 1. INTRODUCTION

Most of the meteoroids impacting the Earth with masses from $2·10^{-8}$ to $2·10^{3}$ kg are of sporadic origin and so the specific parent of the vast majority of these bodies remains unknown (Ceplecha 2001, Jopek & Williams 2013). The identification of the parent bodies of large meteoroids that produce fireballs when they encounter the Earth's atmosphere is of particular interest. In fact, the identification of the progenitors of meteoroid streams is crucial because it increases our knowledge on the origin and evolution of these streams, but also allows us to make valuable deductions about chemical and physical properties of the parent and the meteoroid ejection process or break-up of the parent. Meter-sized particles of cometary composition are destroyed at high altitudes (Ceplecha 2001; Madiedo et al. 2013a), but tougher meteoroids can give rise to deep penetrating bolides that, under appropriate circumstances, may produce meteorite falls. Bodies in the 10- meters range can cause damage to buildings and injure people, as the recent Feb. 15$^{th}$, 2013 event over Chelyabinsk, Russia has demonstrated (for scientific details see Brown et al. 2013).

Fireball networks involving multiple observing sites are very important in this field as they can provide sufficient data for a heliocentric orbit of the ablating meteoroid to be determined. The observations can also provide information regarding the strength and composition of the meteoroid. This in turn implies that information on physical processes such as outgassing, violent collisions and breakups that occurred on a specific parent body can be obtained. A recent review about the relationships between meteoroids and their parents can be found in Williams (2011). The SPanish Meteor Network is undertaking such observations and is involved in the identification of the parent bodies of bright bolides (Madiedo and Trigo-Rodríguez 2008; Madiedo et al.

2013b). The standard way of linking meteoroids with a potential parent body is through the similarity of their orbits. According to this approach, the value of the so-called dissimilarity function, which measures the "distance" between the orbit of the meteoroid and that of the potential parent, is calculated. Thus, orbits are deemed to be similar provided this distance is less than a predefined cut-off value. Porubčan et al. (2004) studied the similarity of meteoroid orbits with their parent bodies for a period of at least 5,000 years before a generic association is claimed, a view later supported by Trigo-Rodriguez et al. (2007). One effect of planetary perturbations on a meteoroid orbit is to cause the argument of perihelion to circulate through all values. Thus, bodies in the NEO region suffer this effect in timescales between 5,000 and 10,000 years. The first dissimilarity criterion was proposed by Southworth and Hawkins (1963), who defined the following function for two orbits A and B:

$$D_{SH}^2 = (e_B - e_A)^2 + (q_B - q_A)^2 + \left(2\sin\frac{I_{BA}}{2}\right)^2 + \left(\frac{e_A + e_B}{2}\right)^2 \left(2\sin\frac{\pi_{BA}}{2}\right)^2 \qquad (1)$$

In this equation q and e are, respectively, the perihelion distance expressed in astronomical units and the orbital eccentricity. $I_{BA}$ is the angle between the two orbital planes and $\pi_{BA}$ the difference between the longitudes of perihelia as measured from the intersection of both orbits. Usually, a cut-off value of 0.15 is adopted for $D_{SH}$ (Lindblad 1971a and 1971b). Alternative versions of this dissimilarity function include those proposed by Drummond (1981), Jopek (1993), Valsecchi, Jopek & Froeschlé (1999) and Jenniskens (2008).

This approach has been recently employed by Madiedo et al. (2013b) to propose a link between the Northern χ-Orionids meteoroid stream and the Potentially Hazardous Asteroid 2008XM1. For this purpose, a software package which searches the NeoDys and the MPC databases was employed. Here we again employ this software tool together with the Mercury 6 orbital integrator (Chambers 1999) to identify likely parent bodies of a bright fireball recorded over the south of Spain during the morning twilight (at 7h45m47.8±0.1s local time) of December 27, 2012.

**2. INSTRUMENTATION AND METHODS**

We have employed high-sensitivity CCD video cameras (models 902H and 902H Ultimate from Watec Co.) to image the fireball analyzed here. This event was recorded from two SPMN observing stations whose locations are given in Table 1 that monitor the night sky by means of an array of these devices. Station #1 works in an autonomous way by means of software described in Madiedo & Trigo-Rodríguez (2010) and Madiedo et al. (2010). These cameras generate interlaced imagery according to the PAL video standard, so that video sequences are produced at a rate of 25 frames per second with a resolution of 720x576 pixels. A detailed description of the operation of these systems is given in Madiedo & Trigo-Rodríguez (2008, 2010), Trigo-Rodríguez et al. (2007).

Station #1 has also been carrying out a continuous spectroscopic monitoring since 2009. A holographic diffraction gratings (1000 lines/mm) attached to the objective lens of some of the above-mentioned cameras is used to image the emission spectra resulting from the ablation of meteoroids in the atmosphere. These can record the spectrum of meteor events brighter than magnitude -3 to -4.

For the reduction of the images obtained for the fireball considered in this work, we first deinterlaced the video sequences provided by our cameras. Thus, even and odd fields were separated for each video frame, and a new video file containing these was generated. As this operation implies duplicating the total number of frames, the frame rate in the resulting video was 50 fps. Then, an astrometric measurement was done by hand in order to obtain the plate (x, y) coordinates of the meteor along its apparent path from each station. The astrometric measurements were then introduced in the Amalthea software (Trigo-Rodríguez et al. 2009; Madiedo et al. 2011), which transforms plate coordinates into equatorial coordinates by using the position of reference stars appearing in the images. Typically, we employ around 40 reference stars for this calibration. However, since the fireball analyzed here occurred under twilight conditions, just 8 calibration stars were available. To address this issue, the images were calibrated by measuring the position of reference stars appearing on images taken by the same cameras at a known instant in time before twilight. This calibration provided the values of altitude and azimuth for each pixel in these images, so that the altitude and azimuth of each point along the meteor apparent path was obtained. By taking into account the appearance time of the bolide, the corresponding equatorial coordinates were then inferred.

The Amalthea software employs the method of the intersection of planes to determine the position of the apparent radiant and also to reconstruct the trajectory in the atmosphere of meteors recorded from at least two different observing stations (Ceplecha 1987). In this way, the beginning and terminal heights of the meteor are inferred. From the sequential measurements of the trajectory coordinates for each video frame and the subsequently inferred trajectory length between frames, the velocity of the meteor along its path is obtained. The preatmospheric velocity $V_\infty$ is found from the measured velocities at the earliest part of the meteor trajectory. Once these data are known, the software computes the orbital parameters of the meteoroid by following the procedure described in Ceplecha (1987).

### 3. OBSERVATIONS: atmospheric trajectory and orbit

The fireball shown in Figure 1 (SPMN code 271212) was recorded under twilight conditions in the morning of December 27, 2012, at 6h45m47.8±0.1s UTC while the emission spectrum of the bolide was also obtained at station #1. Several casual witnesses (private citizens) also reported their observations of the event to the website of the SPMN, via an online form that is available for these purposes. These reports did not contain accurate information about the trajectory of the bolide and so they were not used for the orbit determination. We named this fireball "Macael" (after the town that the fireball overflew in Andalusia), as the beginning of its luminous phase was close to the zenith of this town. The absolute magnitude of this event was -9±1. From the analysis of the atmospheric trajectory we deduce that the parent meteoroid entered the atmosphere with an initial velocity $V_\infty$=16.1±0.3 km s$^{-1}$. The luminous phase started at 80.5±0.5 km above the **sea** level, with the apparent radiant located at α=266.5±0.4º, δ=-6.9±0.3º. Unfortunately, as can be see in Figure 1a, the final part of the fireball's trajectory was out the field of view of the camera located at station #1. Also, as can be seen in Figure 1b, the fireball disappeared behind one of the leaves of a palm tree as seen from station #2, when it was located at a height of 47.0±0.5 km so that the terminal point of its atmospheric path was again not observed. However, the fireball was not observed emerging from behind the leaf, and this constrains the final height to be greater that 39.9±0.5 km. In any case, accurate knowledge of the terminal point of the path is not essential in order to determine the radiant and orbital parameters and these are summarized, respectively, in Tables 2 and 3. The heliocentric orbit of this meteoroid is shown in Figure 1d. The data does not match any currently known meteoroid streams.

## 4. DISCUSSION
### 4.1. Light curve: initial mass

Figure 2 shows the light curve (expressed as absolute magnitude vs. time) obtained for the SPMN271212 "Macael" fireball. This has been obtained from the photometric analysis of the images recorded from station #1 in Table 1 (Sierra Nevada). As can be seen, the fireball reached its maximum brightness during the second half of its atmospheric trajectory, a behaviour that is typical of compact (low porosity) meteoroids (Murray et al. 1999; Campbell et al. 2000). The light curve in Figure 2 also shows that a remarkable drop and re-rise of brightness took place around 4 seconds after the beginning of the event. We assume that the initial photometric mass $m_p$ of the parent meteoroid is the same as the total mass lost due to the ablation process between the beginning of the luminous phase and the terminal point of the atmospheric trajectory, that is

$$\mathbf{m_p = 2 \int_{t_b}^{t_e} I_p / (\tau v^2) dt} \qquad (2)$$

where $t_b$ and $t_e$ are, respectively, the times corresponding to the beginning and the end of the luminous phase and $I_p$ is the measured luminosity of the fireball. The luminous efficiency $\tau$ is defined by

$$\log_{10}(\tau) = -13.24 + 0.77 \log_{10}(v) \qquad (3)$$

for velocities v ranging between 12.5 and 17 km s$^{-1}$ (Ceplecha & McCrosky 1976). Using these equations and the observed light-curve, a pre-atmospheric mass $m_p = 3.7 \pm 0.8$ kg was obtained. The diameter of the meteoroid ranges from 14.3 cm, assuming a density of 2.4 g cm$^{-3}$, to 12.5 cm for a density of 3.7 g cm$^{-3}$. However, it must be taken into account that despite the many papers dealing with the different ways of determining the luminous efficiency, this parameter is one of the least conclusively established values in meteor science. Since the uncertainty in the luminous efficiency given by equation (3) is not known, the calculated error bars for $m_p$ should be regarded as a lower limit for the uncertainty in the meteoroid mass.

### 4.2. Emission spectrum

As mentioned already, one of the video spectrographs operating at station #1 recorded the emission spectrum of the SPMN271212 bolide. The relative spectral sensitivity of this device is shown in Figure 3. To identify the chemical elements and multiplets producing the main lines exhibited by this spectrum we have followed the fitting technique described in Trigo-Rodriguez et al. (2003). Each frame was sky-background-substracted and flat-fielded and the signal was calibrated in wavelength by identifying typical lines appearing in meteor spectra, such as those produced by Na I-1 at 588.9 nm, Mg I-2 at 516.8 nm and Ca I-2 at 422.6 nm. The resulting spectrum integrated along all video frames was also corrected by taking into account the spectral efficiency of the spectrograph, is shown in Figure 4. The most significant contributions, which are well above the noise level of the images, have been highlighted in this plot. Most of the lines identified in the spectrum correspond to Fe I, which is typical for meteor spectra. The signal is dominated by the emission from multiplet Na I-1 at 588.9 nm. The Mg I-2 triplet at 516.8 nm was also identified, although its intensity is low. The Ca I-2 line at 422.6 nm can also be noticed. In the ultraviolet, the most important contribution comes from multiplet Fe I-23 at 358.7 nm. In this region the H and K lines of Ca II at, 396.8 nm and 393.4 nm, respectively, were identified. Nevertheless these appear blended with several Fe I lines. In the red region, atmospheric $N_2$ bands can be seen.

Once the different contributions in the spectrum have been identified, we have obtained an insight into the chemical nature of the meteoroid by analyzing the relative intensities of the Na I-1, Mg I-2 and Fe I-15 multiplets (Borovička et al. 2005). It should be emphasized, however, that these intensities are not proportional to the corresponding relative abundances, since to estimate these abundances an appropriate spectral model must be used. Thus, the Na/Mg and Fe/Mg intensity ratios yield 3.7 and 0.9, respectively. Figure 5 shows the relative intensities of Na I-1, Mg I-2 and Fe I-15 on a ternary diagram. The solid curve in this plot corresponds to the expected relative intensity, as a function of meteor velocity, for chondritic meteoroids (Borovička et al.

2005). As can be seen, the point describing the emission spectrum deviates significantly from the value expected on this curve for a meteor velocity of around 16 km s$^{-1}$. So, this analysis suggests that the meteoroid is achondritic.

In addition, we have estimated the relative chemical abundances of the main rocky elements in the meteoroid with respect to Fe by following the technique described in Borovička (1993) and Trigo-Rodríguez et al. (2003). We have obtained Mg/Fe=0.47±0.06, Na/Fe=0.020±0.003, Ca/Fe=0.10±0.02 and Mn/Fe=0.017±0.005. These clearly deviate from the expected chondritic values Mg/Fe=1.23, Na/Fe=0.07, Ca/Fe=0.07 and Mn/Fe=0.05 (Rietmeijer & Nuth 2000) and point towards an iron-rich achondrite as the likely nature of the SPMN271212 meteoroid. Thus, to get an insight into the likely nature of this particle we have obtained the elemental abundances relative to Si by assuming a value of Fe/Si=0.60, which is typical of HED basalts (Warren et al. 2009). The results are summarized in Table 5, where these relative abundances are compared with those found for other bodies in the Solar System, including the different classes of differentiated achondrites. As can be noticed, the derived low Mg/Si and Na/Si ratios are consistent with an achondrite of basaltic nature (Nittler et al. 2004). The relative chemical abundances of the main rocky elements allow us to suggest that the meteoroid mineralogy was probably closer to Pristine Igneous Eucrites (PIE) or Polymict Eucrites (PE). On the other hand, the Al/Si ratio points to a Basaltic Shergottite, but we should be careful with this appreciation. The reason to be more conservative here is the fact that for the Al/Si ratio, as also for the remarkably low Ca/Si ratio, the inferred abundances should be taken with caution as they are probably lower limits. The reason is that both values are typically underestimated in meteor spectra as a consequence of the inefficiency, particularly for low geocentric velocity meteoroids like the studied here, to bring all available Ca or Al to the vapour phase. In fact, it has been noted (Borovička 1993, 1994; Trigo-Rodríguez et al. 2003) that both elements are associated with refractory minerals that are not fully evaporated during atmospheric ablation. This is why the contribution of Ca and Al in meteor spectra increases for meteoroids with higher geocentric velocity (Trigo-Rodríguez 2002; Trigo-Rodríguez et al. 2003, 2004; Trigo-Rodríguez and Llorca 2007). It is likely that some of these mineral phases might survive the bolide phase as dust, being later on recovered as micrometeorites. Consequently, from the elemental data shown in Table 5, we infer that the SPMN271212 meteoroid had a bulk chemistry similar to that expected for an eucrite or a basaltic shergottite.

**4.3. Orbital analysis: parent body searches**

The ORAS software, which is described in detail in Madiedo et al. (2013b), searches through the NeoDys (http://newton.dm.unipi.it/neodys/) and the Minor Planet Center (http://www.minorplanetcenter.org/iau/mpc.html) databases in order to establish a potential similarity between the orbit of SPMN271212 meteoroid and the orbits of other bodies in the Solar System using the dissimilarity criterion proposed by Southworth & Hawkins (1963). Seven possible candidates were found satisfying the condition $D_{SH}$<0.15 and details are given in Table 4. As can be seen the orbital elements of all of them are quite similar to those of SPMN271212.

By far the lowest value of $D_{SH}$, namely 0.08, was obtained for NEO 2012XJ112. A numerical integration backwards in time for 100,000 years of the orbits of the fireball and this NEO was performed using the Mercury 6 software (Chambers 1999) in order to determine whether their orbital evolution was similar. The gravitational fields of Venus, Earth, Moon, Mars, Jupiter and Saturn were included. As can be seen in Figure 6, the $D_{SH}$ criterion remains below or equal to 0.15 during most of that period, strongly suggesting a link between the fireball and 2012XJ112. In fact, as can be seen in Figure 7, the evolution over time of the orbital elements of 2012XJ112 and the meteoroid is very similar. In order to further test the hypothesis that the asteroid and the SPMN271212 fireball are related, a set of 100 clones was created around the orbit of 2012XJ112. The orbital elements of the clones were spread within the error bars provided by the JPL database (http://ssd.jpl.nasa.gov/) at the 3-sigma confidence interval. The backward orbital evolution of these clones was then obtained in a similar manner to those of 2012XJ112 and SPMN271212. The evolution with time of the $D_{SH}$ criterion for these clones and the fireball was also calculated, the results being shown in Figure 8. We see that $D_{SH}$ remains below 0.15 for all of the clones not just over a 5,000 years time scale required by Porubčan et al. (2004), but also over at least 20,000 years. Further, for 88 out of the 100 clones $D_{SH}$ remains below 0.15 for over 30,000 years while 39 clones fulfill this condition for over 50,000 years and for 16 of clones, $D_{SH}$ remains below 0.15 for these whole integration interval of 100,000 years.  This confirms the potential link between the asteroid and the fireball. Thus, our analysis shows that at present the best fit to the bolide is NEO 2012XJ112. However, there could of course be an as yet undiscovered NEO that will prove to be a better fit as was the case for the Northern χ-Orionids, where initially 2002XM35 was thought to be the parent (Porubčan et al. 2004) but after discovery 2008XM1 proved to be a closer fit (Madiedo et al. 2013b). Unfortunately there are no spectrophotometric data for asteroid 2012XJ112, so it is impossible to check its achondritic nature (R. Binzel, Pers. Comm.).

The fact that the Macael fireball shows deep penetration in the atmosphere together with the particle compactness inferred from the light curve also suggest that its parent is an asteroid and is a potential source of meteorites. Evidence of potential meteorite-dropping bolides linked to the NEO population was previously found by (Trigo-Rodriguez et al. 2007) for asteroid 2002NY40.

Different physical processes could explain the origin of the meteoroid that we recorded as the "Macael" fireball. Thus, for instance, one option would be that the meteoroid was ejected by a fast rotator. An alternative pathway is a gravitational-induced disruption during a close approach of the progenitor asteroid to other planetary bodies, such as the Earth, the Moon, Venus or Mars. This, however, is an interesting case to be checked by dynamicists as it is not the main goal of this paper.

## 5. CONCLUSIONS

We have calculated the atmospheric trajectory, radiant and orbit for a double-station mag. -9±1 fireball recorded by SPMN. The main conclusions derived from the analysis of this bolide, which is listed in our database with code SPMN271212 "Macael", are summarized below:

1) The calculation of the atmospheric trajectory, radiant and orbit revealed that this event did not match any currently known meteoroid stream. The parent meteoroid, with an estimated initial mass of about 3.7 kg, could have penetrated the atmosphere down to a height of about 40 km above the **sea** level.
2) Seven potential parent bodies were suggested on the basis of orbital similarity. The Near Earth Object 2012XJ112 provided the lowest value for the Southworth and Hawkins dissimilarity function, $D_{SH}$ (Southworth and Hawkins 1963).

3) A set of 100 clones around the orbit of asteroid 2012XJ112 was generated. These were integrated backwards in time with the Mercury 6 software during a period of 100,000 years. All of the clones provided $D_{SH} \leq 0.15$ over a time period of over 20,000 years. This confirms the potential link between the SPMN271212 "Macael" fireball and the Apollo asteroid 2012XJ112. The deep penetration in the atmosphere exhibited by the fireball and the compactness of the meteoroid inferred from the analysis of the light curve suggest that 2012XJ112 might be a potential source of meteorites, though none have been found to date.

4) The analysis of the emission spectrum of the SPMN271212 bolide allows us to predict a basaltic (achondritic) nature for the meteoroid.

5) Thus, our analysis also implies that NEO 2012XJ112 is a source of achondrites reaching the Earth. Unfortunately there are no spectrophotometric data on this NEO, but our results could encourage other groups to obtain them.

**ACKNOWLEDGEMENTS**

We acknowledge support from the Spanish Ministry of Science and Innovation (projects AYA2009-13227, AYA2011-26522 and AYA2009-06330-E) and Junta de Andalucía (project P09-FQM-4555). The authors are also grateful to Paul Warren (IGPP/UCLA), Rick Binzel (MIT), Hap McSween (Univ. Tennessee) and Larry Nittler (Carnegie Inst.) for providing useful comments and the bulk chemistry data of achondrite groups.

**REFERENCES**

Borovička J., 1993, A&A, 279, 627.

Borovička J., 1994, Astronomy and Astrophysics Supplement Series, 103, 83.

Borovička J., Koten P., Spurny P., Boček J., Stork R., 2005, Icarus, 174, 15.

Brown P. et al., 2013, Nature, 503, 238.

Campbell M.D. et al., 2000, Meteorit. & Planet. Sci., 35, 1259.

Ceplecha Z., 1987, Bull. Astron. Inst. Cz., 38, 222.

Ceplecha Z., 2001, Collisional processes in the Solar System. Kluwer Academic Publishers, 35.

Ceplecha Z. and McCrosky R.E., 1976, J. Geophys. Res., 81, 6257.

Chambers J.E., 1999, MNRAS, 304, 793.

Drummond J.D., 1981, Icarus 45, 545.

Jenniskens P., 2008, Icarus, 194, 13.

Jopek T.J., 1993, Icarus, 106, 603.

Jopek T.J., Williams I.P., 2013, MNRAS, 430, 2371.

Lindblad B.A., 1971a, Smiths. Contr. Astrophys., 12, 1.

Lindblad B.A., 1971b, Smiths. Contr. Astrophys., 12, 14.

Madiedo J.M., Trigo-Rodríguez J.M., 2008, Earth Moon and Planets, 102, 133.

Madiedo J.M., Trigo-Rodriguez J.M., 2010, 41st Lunar and Planetary Science Conference, abstract # 1504.

Madiedo J.M., Trigo-Rodríguez J.M., Lyytinen E., 2011, NASA/CP-2011-216469, 330.

Madiedo J.M., Trigo-Rodríguez J.M., Ortiz J.L., Morales N., 2010, Advances in Astronomy, 2010, 1.

Madiedo J.M. et al., 2013a, Icarus, in press.

Madiedo J.M., Trigo-Rodríguez J.M., Williams I.P., Ortiz J.L., Cabrera J., 2013b, MNRAS, 431, 2464-2470.

Murray I.S., Hawkes R.L., Jenniskens P., 1999, Meteorit. & Planet. Sci., 34, 949.

Nittler L., McCoy T.J., Clark P.E., Murphy M.E., Trombka J.I., Jarosewich E., 2004, Antarct. Meteorite Res., 17, 233.

Porubčan V., Williams I.P., Kornoš L., 2004, Earth, Moon and Planets, 95, 697.

Rietmeijer F., 2002, Chemie der Erde, 62, 1.

Rietmeijer F. and Nuth J.A., 2000, Earth, Moon and Planets, 82-83, 325.

Southworth R.B., Hawkins G.S., 1963, Smithson Contr. Astrophys., 7, 261.

Trigo-Rodríguez J.M., 2002, Ph.D. Thesis, Universitat de Valencia.

Trigo-Rodríguez J.M. and Llorca J., 2007, Adv. Space Res. 39, 517.

Trigo-Rodríguez J.M., Llorca J., Borovička J., Fabregat J., 2003, Meteorit. Planet. Sci., 38, 1283.

Trigo-Rodríguez J.M., Madiedo J.M., Williams I.P., et al., 2009, MNRAS, 394, 569.

Trigo-Rodríguez J.M. et al., 2004, MNRAS, 348, 802.

Trigo-Rodríguez J.M. et al., 2007, MNRAS 382, 1933.

Valsecchi G., Jopek T., Froeschlé C., 1999, MNRAS, 304, 743.

Warren P.H., Kallemeyn G.W., Huber H., Ulff-Møller F., Choe W., 2009, GCA, 73, 5918.

Williams I.P., 2011, A&G, 52, 2.20

**TABLES**

Table 1. Geographical coordinates of the meteor observing stations involved in this work.

| Station # | Station name | Longitude (W) | Latitude (N) | Alt. (m) |
|---|---|---|---|---|
| 1 | Sierra Nevada | 3º 23´ 05" | 37º 03´ 51" | 2896 |
| 2 | La Murta | 1º 09' 50" | 38º 05' 54" | 94 |

Table 2. Trajectory and radiant data for the SPMN271212 fireball (J2000). $H_b$ and $H_e$ are the beginning and ending height, respectively. $\alpha_g$ and $\delta_g$ denote the equatorial coordinates of the geocentric radiant. $V_\infty$, $V_g$ and $V_h$ are the initial (preatmospheric) velocity, the geocentric velocity and heliocentric velocity of the meteoroid, respectively. $H_e$ is constrained to the values shown in the table, as explained in the text.

| $H_b$ (km) | $H_e$ (km) | $\alpha_g$ (º) | $\delta_g$ (º) | $V_\infty$ (km s$^{-1}$) | $V_g$ (km s$^{-1}$) | $V_h$ (km s$^{-1}$) |
|---|---|---|---|---|---|---|
| 80.5±0.5 | <47.0±0.5 >39.9±0.5 | 277.3±0.4 | -16.8±0.3 | 16.1±0.3 | 11.2±0.3 | 32.5±0.3 |

Table 3. Orbital parameters (J2000) for the SPMN271212 fireball.

| a (AU) | e | i (º) | ω (º) | Ω (º) |
|---|---|---|---|---|
| 1.187±0.008 | 0.37±0.01 | 2.3±0.3 | 95±1 | 275.7841±10$^{-4}$ |

Table 4. Potential parent bodies found for the Macael fireball on the basis of the Southworth and Hawkins dissimilarity criterion. Orbital elements were taken from the JPL Small-Body Database (http://ssd.jpl.nasa.gov/).

| Object name | $D_{SH}$ | a (AU) | e | i (º) | ω (º) | Ω (º) |
|---|---|---|---|---|---|---|
| 1998HL3 | 0.14 | 1.1285 | 0.3659 | 2.6788 | 188.0885 | 163.7185 |
| 2003MA3 | 0.12 | 1.1053 | 0.4021 | 1.4127 | 228.9189 | 152.6523 |
| 2000AP138 | 0.14 | 2.3324 | 0.1300 | 9.6802 | 52.2199 | 99.0709 |
| 2005XA8 | 0.11 | 1.4181 | 0.4368 | 5.3470 | 111.2010 | 254.1692 |
| 2011FR29 | 0.15 | 1.0575 | 0.3465 | 3.6196 | 175.8290 | 215.3855 |
| 2012BK14 | 0.14 | 0.9824 | 0.1922 | 1.5059 | 254.3142 | 118.7667 |
| 2012XJ112 | 0.08 | 1.2220 | 0.3695 | 2.4347 | 102.3764 | 256.9244 |

Table 5. Computed elemental abundances relative to Si of SPMN271212 meteoroid compared with the inferred for other materials in the Solar System (Rietmeijer & Nuth 2000; Rietmeijer 2002; Trigo-Rodríguez 2002; Warren et al. 2009). The number of HED meteorites averaged in Warren et al. data are 29 PID, 35 PIE, 34 PE, and 40 How. Martian rocks and SNC meteorite values were provided by McSween (pers. comm.). The CHA averaged value is from Chassigny and NWA2737.

| Material | Mg/Si | Na/Si | Ca/Si | Mn/Si |
|---|---|---|---|---|
| SPMN271212 | 0.25±0.06 | 0.012±0.003 | 0.06±0.02 | 0.010±0.005 |
| Pristine Igneous Diogenites (PID) | 0.647 | 0.0009 | 0.039 | 0.015 |
| Pristine Igneous Eucrites (PIE) | 0.204 | 0.013 | 0.304 | 0.018 |
| Polymict Eucrites (PE) | 0.225 | 0.013 | 0.288 | 0.017 |
| Howardites (How) | 0.399 | 0.008 | 0.193 | 0.016 |
| Nakhlites (Mars) | 0.34 | (0.05) | 0.32 | - |
| Bounce Rock (Shergottite, Mars) | 0.17 | 0.05 | 0.287 | 0.009 |
| Chassignites (CHA) | 1.19 | 0.006 | 0.031 | 0.021 |
| Martian basaltic breccia (NWA7034) | 0.21 | 0.125 | 0.287 | 0.009 |
| ALH84001 (ortopyroxenite) | 0.61 | 0.004 | 0.053 | 0.014 |
| CI chondrites | 1.06 | 0.060 | 0.071 | 0.051 |
| CM chondrites | 1.04 | 0.035 | 0.072 | 0.046 |

**FIGURES**

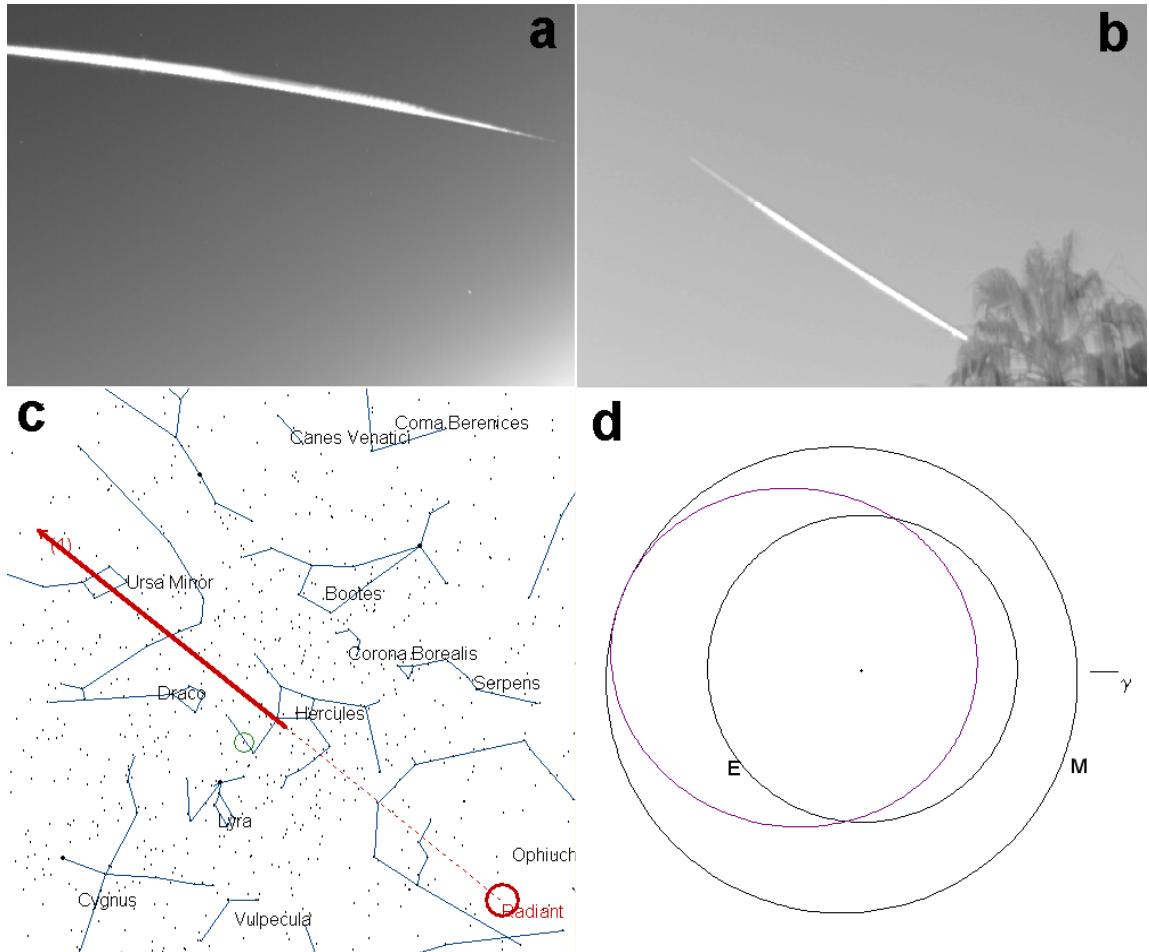

Figure 1. Composite images of the "Macael" fireball (SPMN271212) recorded from Sierra Nevada (a) and Molina de Segura (b). c) Apparent trajectory in the sky as observed from Sierra Nevada. d) Projection on the ecliptic plane of the heliocentric orbit of the parent meteoroid.

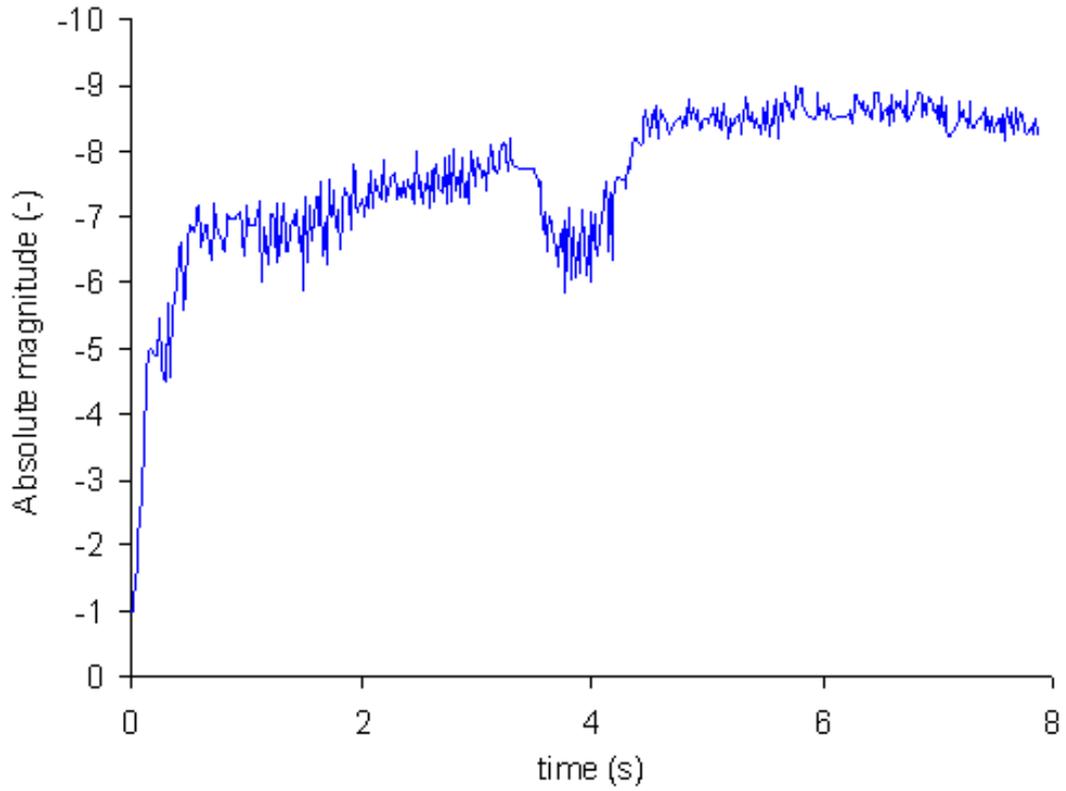

Figure 2. Light curve of the SPMN271212 bolide (absolute magnitude vs. time). This plot shows the sharp drop in brightness discussed in the text that took place around t=4s.

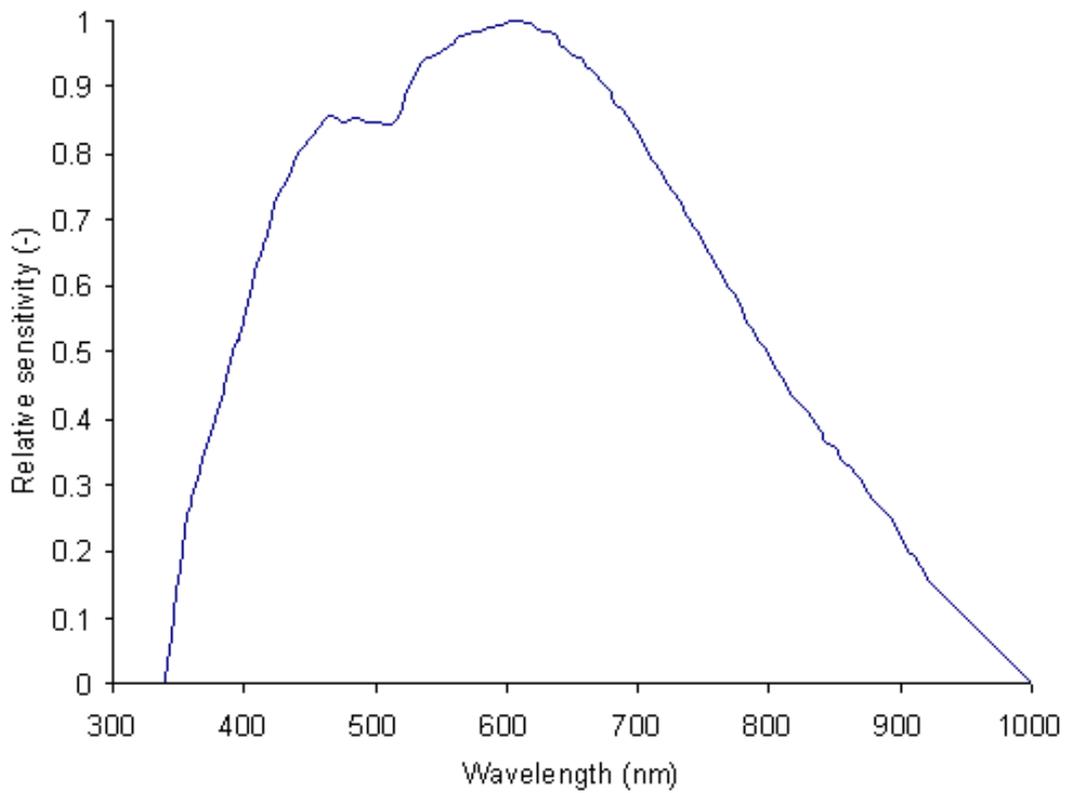

Figure 3. Relative spectral sensitivity of the spectrograph employed to obtain the emission spectrum of the SPMN271212 fireball.

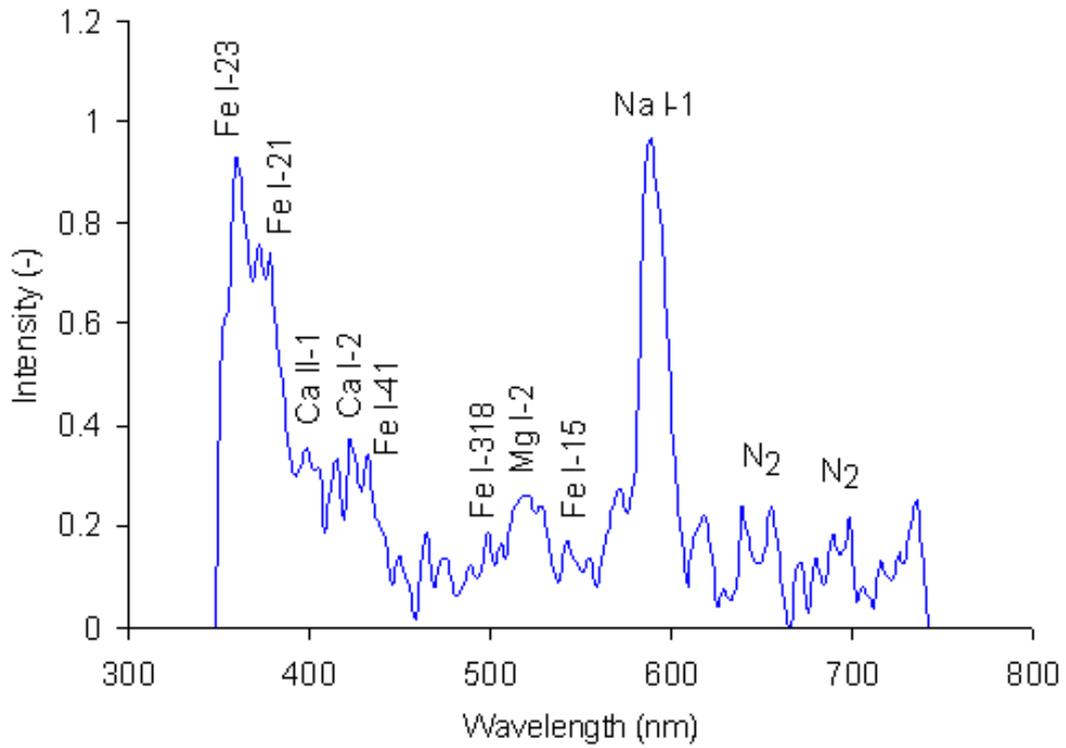

Figure 4. Calibrated emission spectrum of the SPMN271212 fireball. Intensity is expressed in arbitrary units.

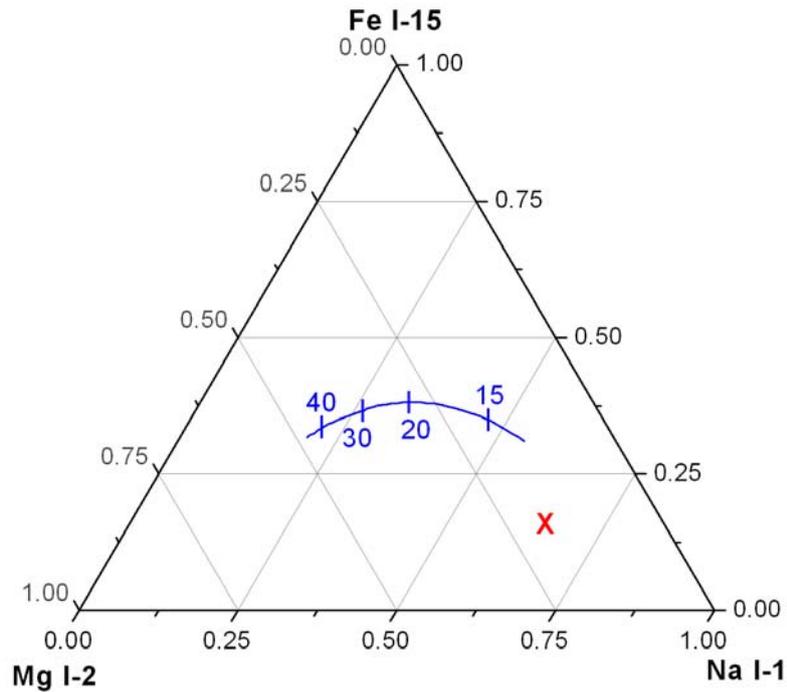

Figure 5. Expected relative intensity (solid line), as a function of meteor velocity (in km s$^{-1}$), of the Na I-1, Mg I-2 and Fe I-15 multiplets for chondritic meteoroids (Borovička et al., 2005). The cross shows the experimental relative intensity obtained for the SPMN271212 fireball.

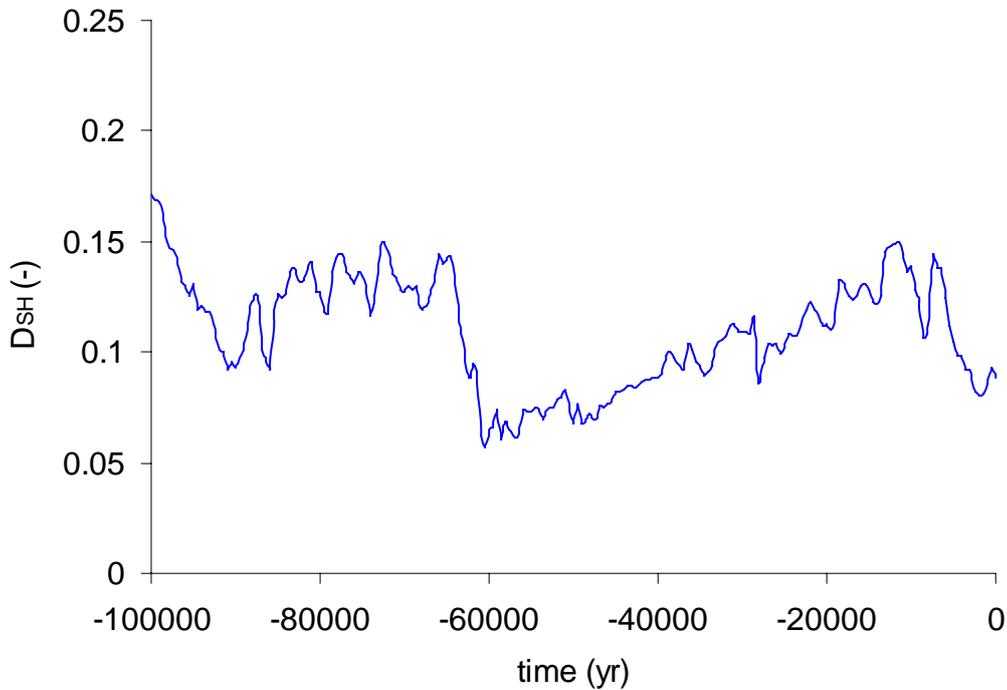

Figure 6. Evolution with time, from the arrival of the meteoroid to Earth, of the $D_{SH}$ criterion for NEO 2012XJ112 and the SPMN271212 fireball.

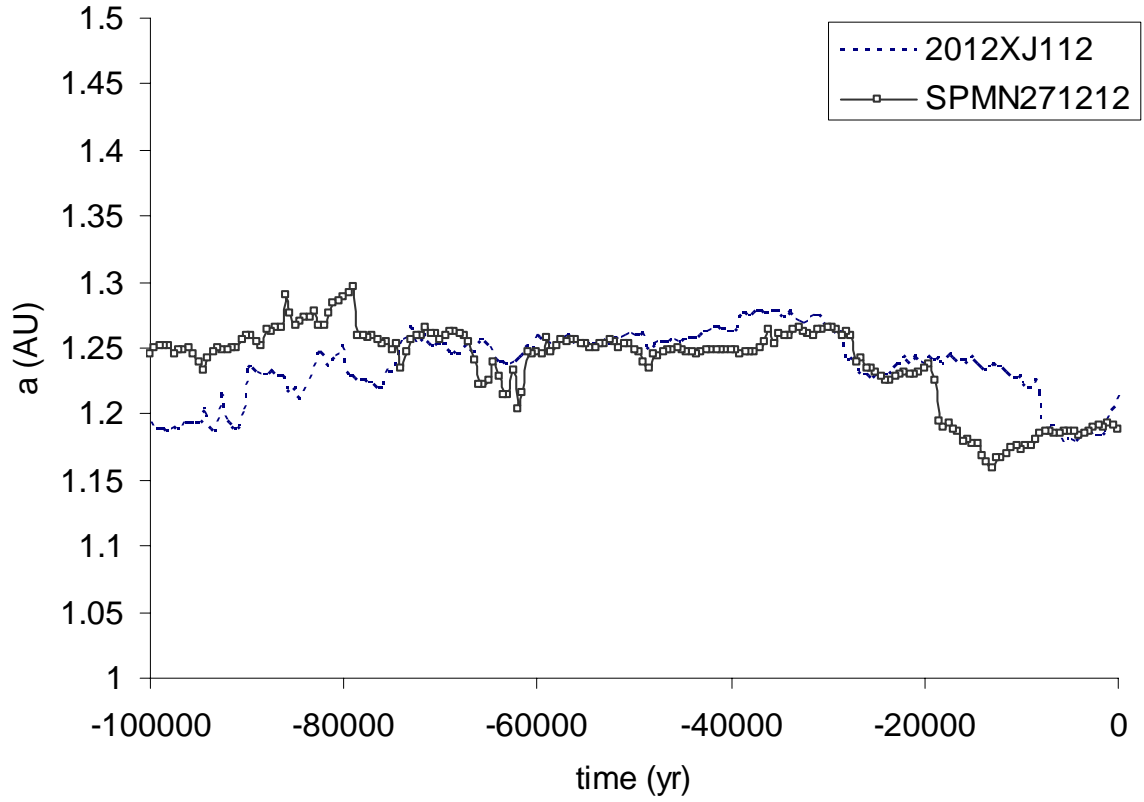

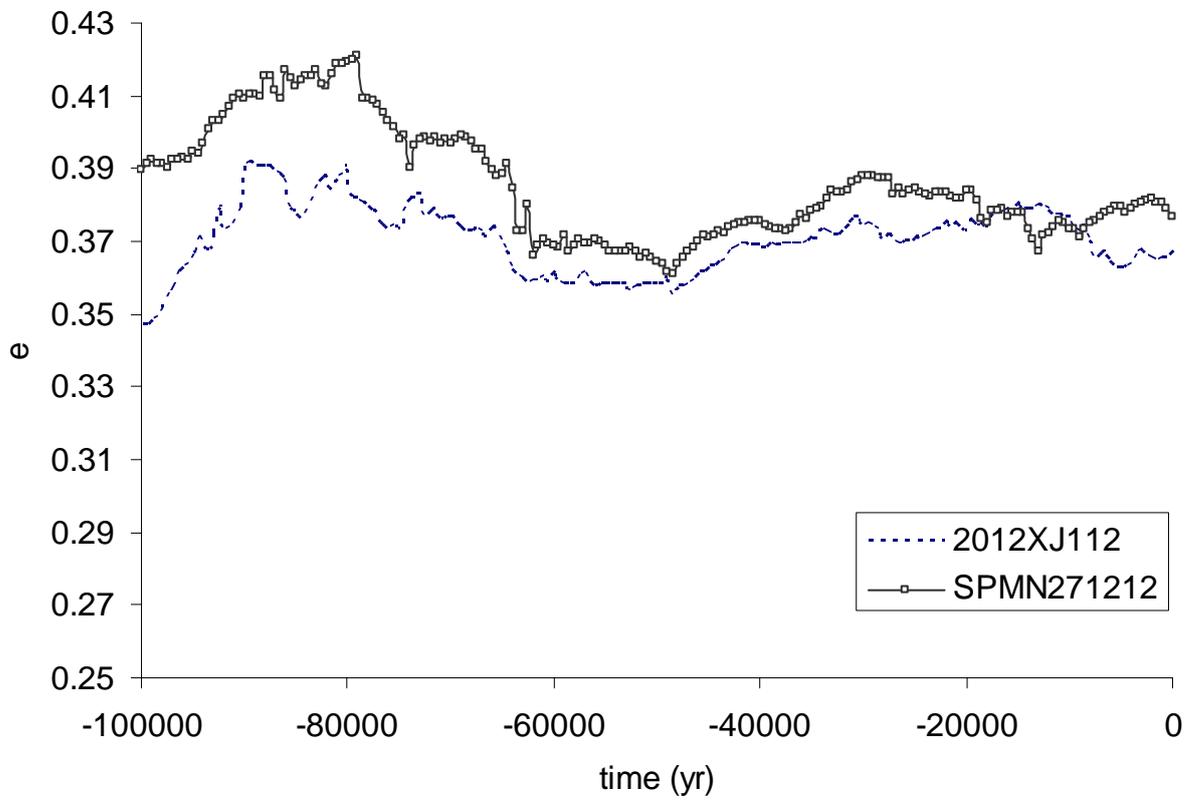

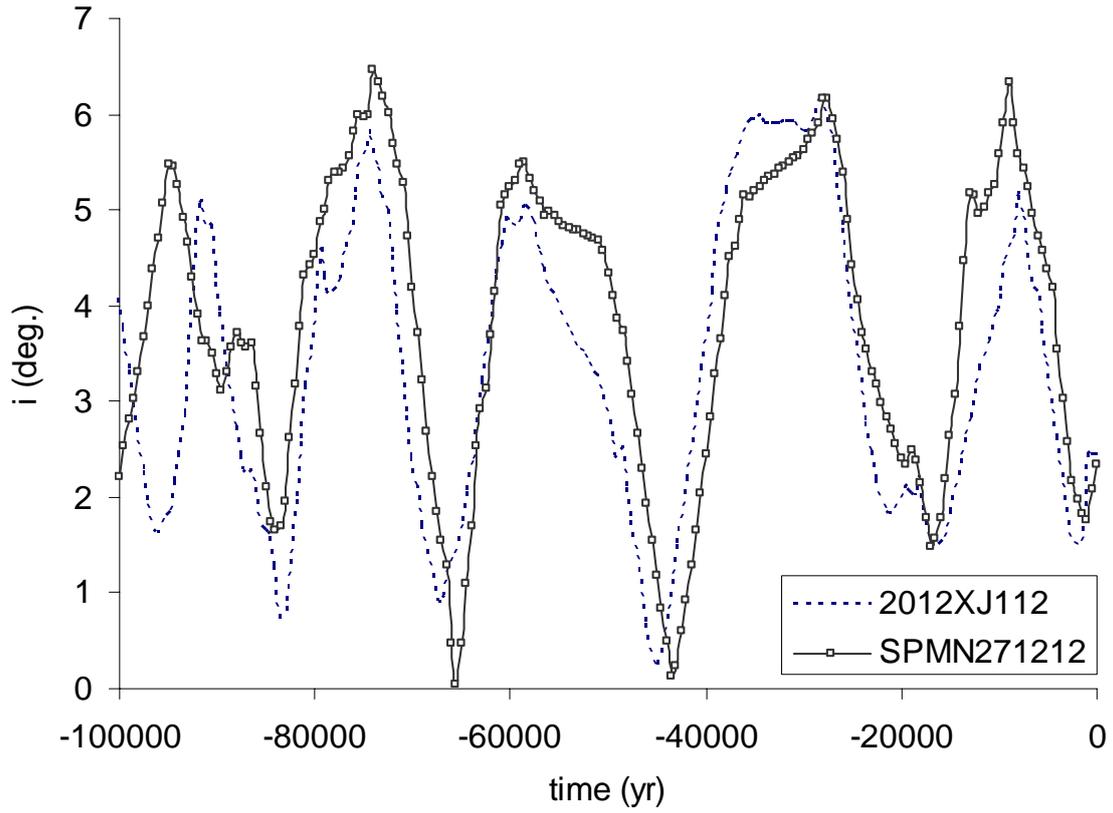

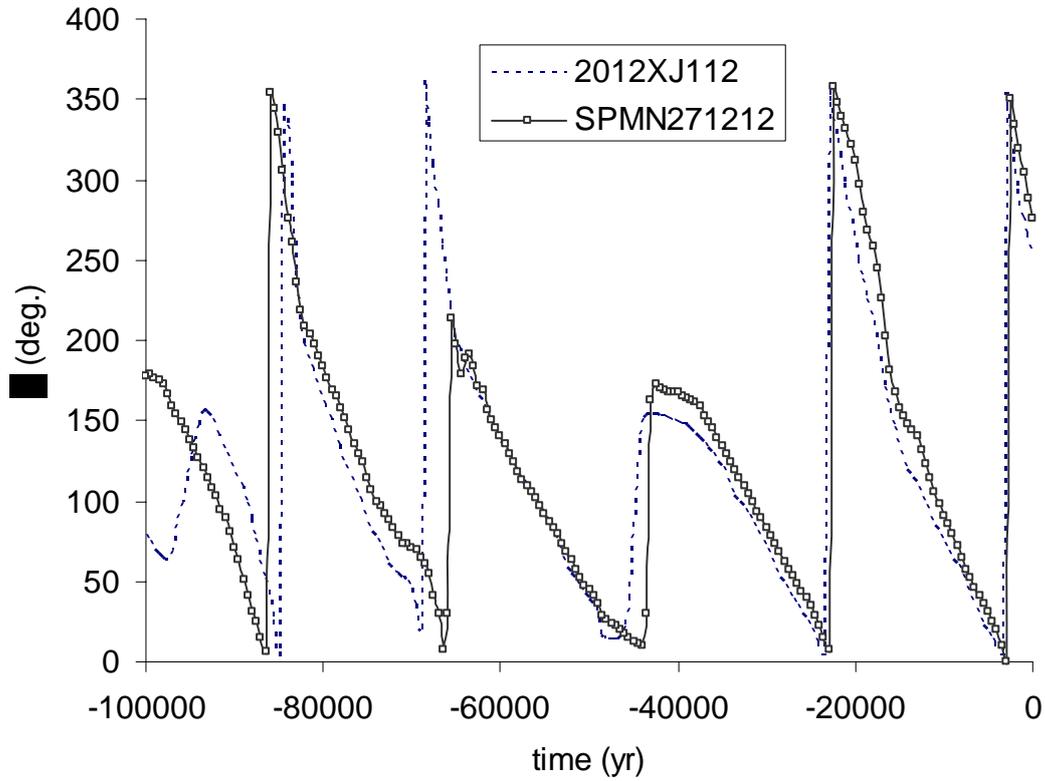

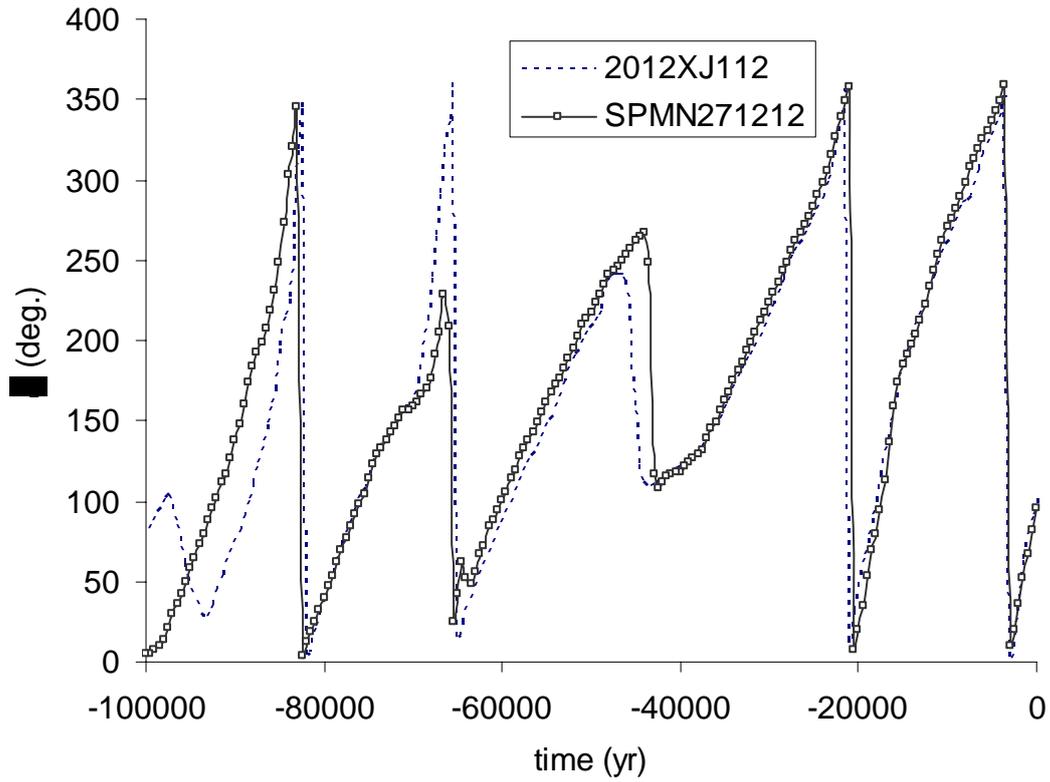

Figure 7. Evolution with time, from the arrival of the meteoroid to Earth, of the orbital elements of NEO 2012XJ112 and the SPMN271212 fireball, where a is the semi-major axis, e the orbital eccentricity, i the inclination, Ω the longitude of the ascending node and ω the argument of perihelion.

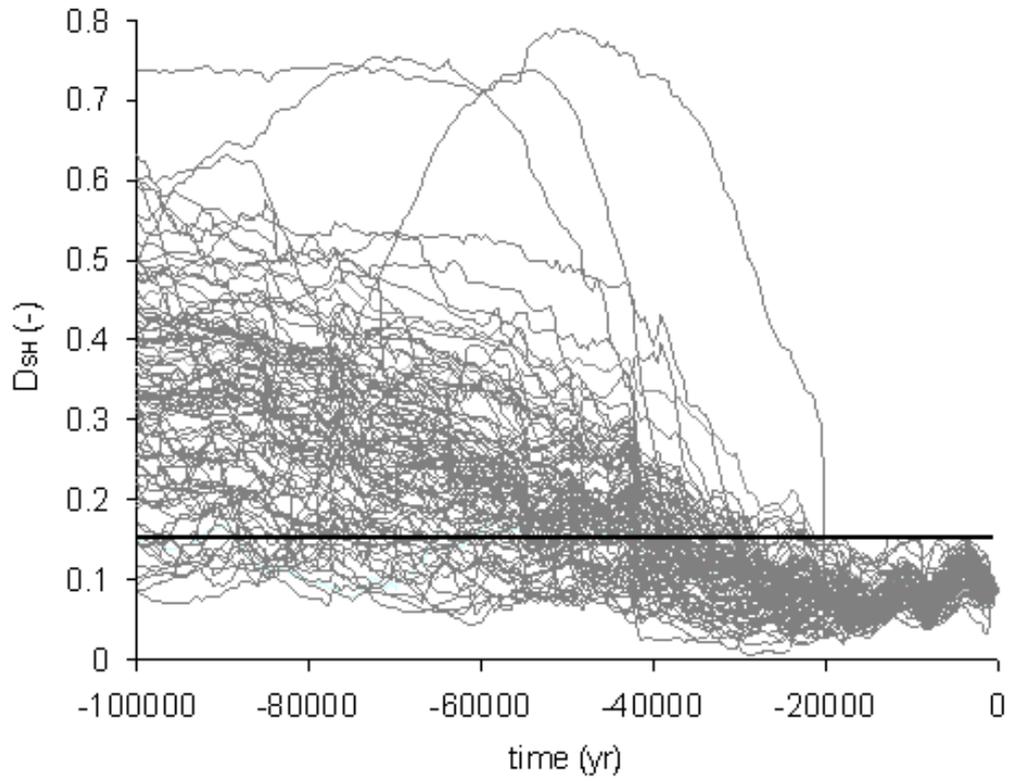

Figure 8. Evolution with time, from the arrival of the meteoroid to Earth, of the $D_{SH}$ criterion for the SPMN271212 fireball and 100 clones of NEO 2012XJ112. The horizontal solid line shows the cut-off value for the dissimilarity function ($D_{SH}$=0.15).